\documentstyle[sprocl]{article}

\def\Journal#1#2#3#4{{#1} {\bf #2}, #3 (#4)}

\def\NPB{{\em Nucl. Phys.} B}
\def\PLB{{\em Phys. Lett.}  B}

\def\PRD{{\em Phys. Rev.} D}

\def\JAD{\em Jad. Fiz.}

\def\be{\begin{equation}}
\def\ee{\end{equation}}
\def\bea{\begin{eqnarray}}
\def\eea{\end{eqnarray}}
\newcommand{\beqna}{\begin{eqnarray}}
\newcommand{\eeqna}{\end{eqnarray}}
\newcommand{\gapproxeq}{\lower.7ex\hbox{$\;\stackrel{\textstyle>}{\sim}\;$}}
\newcommand{\lapproxeq}{\lower.7ex\hbox{$\;\stackrel{\textstyle<}{\sim}\;$}} 

\begin{document}


\title{\begin{flushright} \small{hep-ph/9607476} \end{flushright} 
\vspace{0.6cm}
HAVE GLUONIC EXCITATIONS BEEN FOUND?  \footnote{Talk presented at XIV International Conference on Particles and Nuclei (PANIC96) (Williamsburg, 1996), ed. C. Carlson.}
}

\author{PHILIP R. PAGE \footnote{prp@a13.ph.man.ac.uk}}

\address{Department of Physics and Astronomy, University of Manchester, \\ Manchester M13 9PL, UK}

\maketitle\abstracts{New experimental information on the non--exotic $J^{PC}=0^{-+}$
isovector seen at 1.8 GeV by VES yields convincing
evidence of its excited gluonic (hybrid) nature when a critical study
of alternative quarkonium assignments is made in the context of
$^3P_0$ decay by flux--tube breaking. Production of this gluonic excitation via meson
exchange is promising, although its 
two photon production vanishes.}
  
Based on the phenomenological success\cite{geiger94} of the 
$^3P_0$ hadronic decay model, the decay modes of $Q\bar{Q}$ systems with
an explicit gluonic excitation (hybrids) have been predicted\cite{page} 
in a non--relativistic flux--tube model. Hybrids
are predicted to have mass 1.8 -- 1.9 GeV\cite{geiger94}, exactly in the region
where a $J^{PC}=0^{-+}$ isovector resonance has recently
been seen\cite{ves}. The mass of this state also makes it a candidate for 
radial $3^1S_0 \; Q\bar{Q}$ ($\pi_{RR}$). 
The decay of hybrids to ``S+S''--wave mesons
are expected\cite{ves} to vanish for identical mesons, and
to be suppressed proportional to the difference of their ``sizes'' 
\cite{page} for 
non--identical mesons. The dominant decay channel is hence to
``P+S''--wave mesons.  

VES\cite{ves} (and BNL\cite{ves}) detect a prominent resonance at $\sim 1.8$ GeV 
with width $\sim 200$ MeV in the ``P+S''
channels $\pi f_0(980)$,
$\pi f_0 (1300)$, $\eta a_0 (980)$ and
$(K\bar{K}\pi)_S$. On the other hand, the resonance is absent\cite{ves}
in the ``S+S'' channels $\pi\rho$ and $\bar{K}K^*$. There is also
possible\cite{private} evidence for the (weak) mode $\pi f_0(1500)$
where the gluonic excitation de--exites to the gluonium candidate $f_0(1500)$. 
The foregoing clearly supports a hybrid
interpretation. The predicted
widths for a hybrid $\pi_H$ at $\sim 1.8$ GeV are\cite{page}
(in MeV)

\vspace{-0.5cm}

\beqna
\label{flux}
\pi f_0(1300)\sim 170; \; \pi f_2\sim 5; \; \pi\rho
\sim 30; \;  \bar{K}K^*\sim 5; \; \pi\rho_R\sim 30 \nonumber &  & \\  
\hspace{-1cm} K^*\bar{K}^*\sim 0; \; \rho\omega\sim 0; \; 
\eta a_0 \sim 120; \;  \pi f_0\sim 160
\eeqna

\vspace{-0.2cm}

\noindent where the last two modes assume that $a_0,f_0$ are $^3P_0 \; Q\bar{Q}$. 

The widths expected for $\pi_{RR}$ are often distinctively {\it different} \cite{page} from those of hybrids.
{\it (i)} $\pi f_0 (1300)$ is very much {\it suppressed} ($< 10$ MeV over 
parameter space) relative to the prediction for $\pi_H$ (Eq. \ref{flux}) and 
either $\pi_{RR} \rightarrow \pi \rho , \pi f_2$ or $\bar{K}K^*$ whereas the data show that it is much
larger than all of these. 
{\it (ii)} The same is true of $\bar{K} K^*_0 (1430)$,
which is threshold forbidden and manifested as $(K\bar{K}\pi)_S$.
For $\pi_H$ at $2$ GeV $\bar{K} K^*_0$ is substantial at 200 MeV, consistent with the data,
while for
$\pi_{RR}$ it is {\it suppressed} at $0-20$ MeV due to a node in the amplitude.
The strong affinity of $K\bar{K} \rightarrow f_0(980)$ is probably 
responsible\cite{page} for the observed strong\cite{ves} coupling to $\pi f_0(980)$.
{\it (iii)} For $\pi_{RR}$ the $\rho \omega$ channel is expected to be 
{\it prominent} \cite{page} at $0-120$ MeV. In contrast, 
$\rho \omega$ vanishes for $\pi_H$ (Eq. \ref{flux}) independent of the wave functions assumed
in the flux--tube model\cite{page}. The $\rho \omega$ signal\cite{private} 
builds up significantly below 1.8 GeV with width $\sim 300$ MeV, 
although a resonant signal has not yet been established.
It is tempting  to suggest that this indicates the detection of a seperate state, the $\pi_{RR}$,
different from the 1.8 GeV $\pi_{H}$ with width $\sim 200$ MeV. 
{\it (iv)} The possible existence of a seperate state is corroborated by the $\pi f_2$ channel which may 
also be distinctive. For $\pi_H$ $\pi f_2$ is small (Eq. \ref{flux}) 
whereas it is possibly {\it larger} \cite{page} ($0 - 30$ MeV) for $\pi_{RR}$. 
The data\cite{ves} show a small
$\pi f_2$ peak at 1.7 GeV, certainly
below the 1.8 MeV region, though further analysis and data
are required.

$\pi_H$ and $\pi_{RR}$ have in {\it common} that $\pi\rho$ is suppressed
($0-30$ MeV for $\pi_{RR}$ due to a node)
consistent with the data\cite{ves} which show no signal in the 1.7 to 2 GeV 
mass region. We suggest searching for coupling to the $\pi\rho$ channel, and
further determinations of the mass and width of the state seen in 
$f_2 \pi$ and $\rho\omega$.

At both VES and BNL the $0^{-+}$ was produced in $\pi^- N \rightarrow
0^{-+} N$ at high energy via either diffractive or $\rho$ exchange. 
In the case of $\rho$ exchange the width corresponding to the
$\pi\rho$ vertex of $\pi_H$ is bounded above\cite{page} by 150 MeV, and is 
expected to be $\gapproxeq 20\%$ of this
value (see Eq. \ref{flux}) since the $\rho$ is off--shell 
and hence of potentially very different
``size'' than the on--shell $\pi$. This may lead to  
significant production of $\pi_H$ in photoproduction
on nuclei through $\pi$ exchange, with the photon 
coupling to $\rho$ (with upper bound 270 keV\cite{page}); and would be especially 
significant at low energy facilities like an upgraded 
CEBAF where $\pi$ exchange would be dominant.

An unfortunate corollary of the lack of coupling of $\pi_H$ to $\rho\omega$
mentioned before, is that when the $\rho$ and $\omega$ couple to photons, the
two photon width and production of $\pi_H$ vanish. 
In addition,  the photoproduction of $\pi_H$ via $\rho$ or $\omega$
exchange vanishes\cite{page}.
Photon coupling via intermediate vector mesons is currently the only way of effecting 
flux--tube model photonic couplings for $\pi_H$.

\end{document}